# Scalable Triadic Analysis of Large-Scale Graphs: Multi-Core vs. Multi-Processor vs. Multi-Threaded Shared Memory Architectures

George Chin Jr., Andres Marquez, Sutanay Choudhury and John Feo
*Pacific Northwest National Laboratory*
*{George.Chin, Andres.Marquez, Sutanay.Choudhury, John.Feo}@pnl.gov*

## Abstract

*Triadic analysis encompasses a useful set of graph mining methods that are centered on the concept of a triad, which is a subgraph of three nodes. Such methods are often applied in the social sciences as well as many other diverse fields. Triadic methods commonly operate on a triad census that counts the number of triads of every possible edge configuration in a graph. Like other graph algorithms, triadic census algorithms do not scale well when graphs reach tens of millions to billions of nodes. To enable the triadic analysis of large-scale graphs, we developed and optimized a triad census algorithm to efficiently execute on shared memory architectures. We then conducted performance evaluations of the parallel triad census algorithm on three specific systems: Cray XMT, HP Superdome, and AMD multi-core NUMA machine. These three systems have shared memory architectures but with markedly different hardware capabilities to manage parallelism.*

## 1. Introduction

Large-scale graphs arise in many different fields, domains, and problem areas such as biology, economics, sociology, and cybersecurity. Many of these graphs are “scale-free” [1]. Unlike random graphs that have evenly distributed edges among nodes, a scale-free graph has an edge distribution that follows a power law. A scale-free graph will generally have a limited number of nodes with high degrees while the large majority of nodes will have very low degrees. Scale-free graphs are abundant in the real-world in various kinds of networks such as the World Wide Web, electric power grids, social networks, citation networks, and biological networks.

A common analysis applied on graphs is the mining of frequent subgraphs. In social sciences and related fields, *triads* or subgraphs of three nodes are of particular importance, where they serve as an analytical foundation for understanding larger social groups [2]. Decades of social science research have focused on examining triad distributions in social networks to explain social behaviors and tendencies.

The analysis and mining of large graphs using techniques or algorithms such as triadic analysis have been found to be extremely challenging, because many graph mining algorithms do not scale well when graphs reach tens of millions to billions of nodes and edges [3, 4]. Often, the solution to analyzing large graphs is to develop approximate graph mining methods [5] and/or to optimize the graph mining algorithms to execute on high-performance computers [3, 6, 7].

Analyzing large, complex graphs on high-performance computers presents a unique set of challenges. Graph data are often widely dispersed in memory. This leads to poor data access patterns that are unable to make good use of memory hierarchies. Since data access patterns depend on the actual data, the usefulness of prefetching techniques is also limited. The amount of processing that occurs for each node is often very low, so each memory access may follow with very little actual computation. Scale-free graphs pose additional challenges for high-performance computers, because their power-law edge distributions may cause the amount of memory required and work performed to vary widely from one graph node to another. Based on such factors, “graph computations often achieve a low percentage of theoretical peak performance on traditional processors [7].”

In this paper, we describe how we optimized and parallelized a triad census algorithm to take advantage of the capabilities of shared memory architectures. We also compare the performance of the parallel triad census algorithm across three shared memory systems: Cray XMT, HP (Hewlett-Packard) Superdome, and AMD (Advanced Micro Device) multi-core NUMA (Non-Uniform Memory Access*)* machine.

## 2. Cray XMT

Parallel triad census algorithm development was performed on a Cray XMT machine, but the optimizations are generally applicable to any shared memory architecture. In previous work [8], we implemented and evaluated three preliminary triad census algorithms on the Cray XMT. In this paper, we present our latest, most scalable parallel triad census implementation and evaluate it using scale-free graphs on three different shared memory systems.

The Cray XMT is a distributed shared-memory architecture based on the Cray XT platform, including its high-speed interconnect and network 3D-Torus topology, as well as service and I/O nodes. The compute nodes of the XMT utilize a configuration based on four multithreaded processors known as "Threadstorm" processors. Each Threadstorm processor maintains 128 hardware threads (streams) and their associated contexts. Assuming all data- and resource-dependencies are met, a stream can be scheduled for instruction issue in a single cycle. Each instruction can hold up to three operations, comprising a control, arithmetic, and memory operation.

Rather than focusing on reducing the latency of a single access to memory, the multithreaded Cray XMT processor is designed to tolerate the latency of memory references. On a cycle by cycle basis, each processor context switches among 128 independent hardware instruction streams so that while some streams are waiting for memory operations to complete, other streams may continue to execute. Furthermore, each instruction stream may have up to eight memory references outstanding at a time. In this way, the Cray XMT can continue to make progress on large parallel tasks, effectively hiding the latency to memory.

In addition, the Cray XMT supports the virtualization of threads, where a program may generate massive numbers of software threads that are mapped onto the physical processors. The XMT provides support for fast, dynamic thread creation and destruction along with low-cost scheduling for thousands of concurrent actors. These attributes facilitates dynamic load-balancing and allows applications to adapt the number of concurrent threads to the size of the data or complexity of the problem.

The XMT also supports word-level synchronization, whereby each word of memory may be independent locked, thus reducing potential memory contention issues that would likely ensue with the large numbers of available threads.

The XMT's shared address space and fine-grain thread management techniques make the machine an ideal candidate to execute graph algorithms on large and scale-free graphs. Various graph applications have been implemented and evaluated on the Cray XMT and its predecessor MTA-2 systems including power system state estimation [9], partial dimension trees [10], and Boolean satisfiability solvers [11].

Most of the code development and performance testing was conducted on a 128-processor, 1TB shared memory XMT system located at the Pacific Northwest National Laboratory (PNNL). This system is equipped with 500 MHz Threadstorm 3.X processors. Each processor is attached to 8GB of DDR1 memory. Network capability between the 32 compute blades is provided by a 3D-Torus Seastar-2 interconnect that exhibits a round trip latency of 1.8 µs. We also had limited access to a 512-processor, 4 TB shared memory XMT system housed in Cray's development laboratory in Chippewa Falls, Wisconsin. This larger XMT system is equipped with Threadstorm 3.0.X pre-production processors also running at 500MHz.

## 3. Social networks and triadic analysis

In social network analysis, triadic methods in social network analysis focus on the concept of a *triad*. A triad is a subgraph of three actors and the directed relationships among them. A directed graph has exactly $n(n-1)(n-2)/6$ triads including null triads, where $n$ equals the number of nodes. Triadic methods are considered local methods in the sense they separately examine the properties of subsets of actors as opposed to global methods that simultaneously examine the properties of all actors in the network.

A triad has 64 possible states based on the existence of directed edges among the three actors. Triad states indicate important properties about the triad. For example, the triad states shown in Fig. 1 exhibit properties of reciprocity, transitivity, and intransitivity. As shown in Fig. 2, we can build a *triad census* by capturing the frequencies in which the triads of a network fall into one of the 64 possible triad states. We may condense a 64-element triad census down to a 16-element triad census by considering isomorphic cases, where certain triad types are structurally-equivalent and may directly map onto one another. Computing the triad census is commonly the most computationally-expensive part of triadic methods.

In an application, we applied the triad census algorithm to detect anomalies and potential threats in computer network traffic. Fig. 3 presents four different computer network activities that a computer security analyst may wish to monitor and the corresponding triads that are relevant to each of those patterns. By computing the triad census of a computer network at fixed time intervals, we can track the proportions of triad types relative to one another as well as how these proportions change over time. As shown in Fig. 4, we have developed a monitoring tool that will identify when specific triad patterns are occurring in the network traffic outside their normal behavior and notify computer security analysts when specific combinations of triads signifying a potential security threat or anomaly is detected.

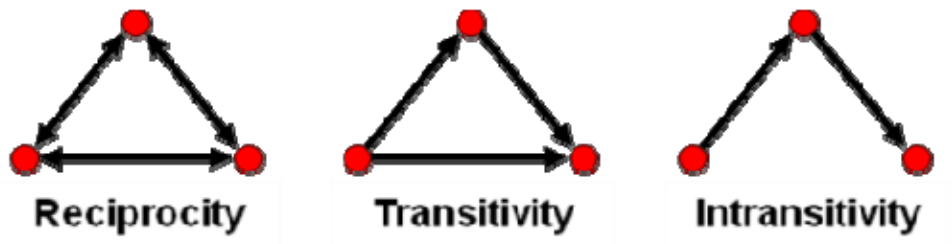

**Figure 1. Triads with specific graph properties.**

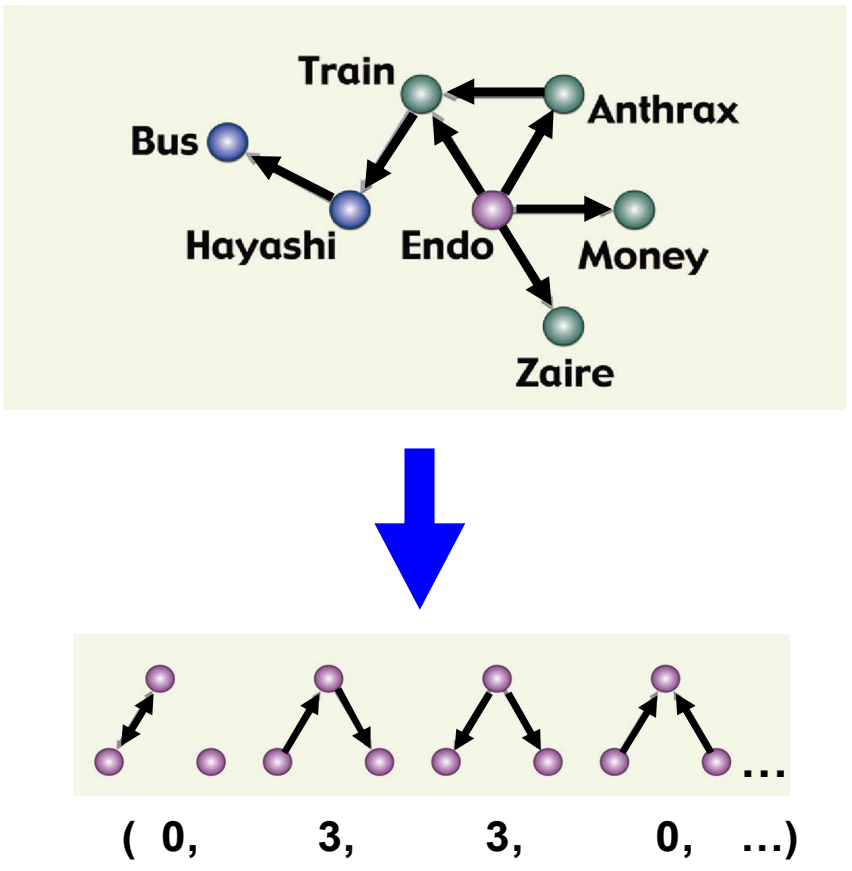

**Figure 2. Creation of a triad census.**

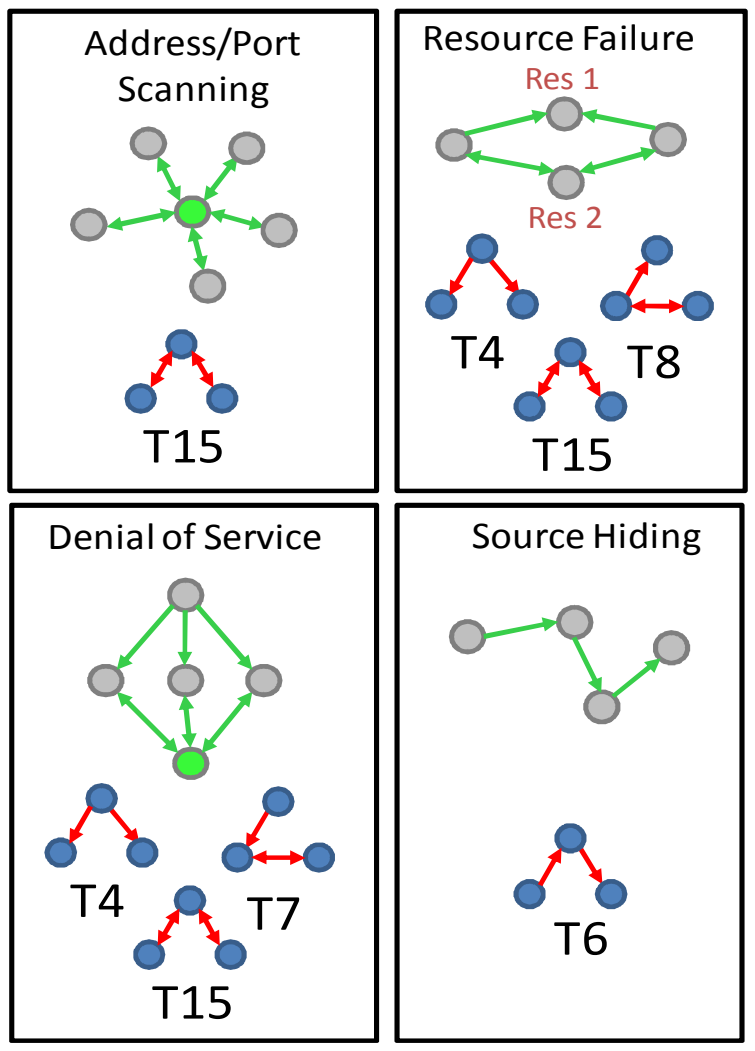

**Figure 3. Computer network security threat and anomaly triad patterns.**

## 4. Triadic census algorithm

The simple, naive algorithm for computing the triad census would visit every possible combination of three nodes in the graph, which would result in a computational complexity of $O(n^3)$, where $n$ equals the number of nodes. Moody [12] developed a more efficient triad census algorithm with a computational complexity of $O(n^2)$, where $n$ equals the number of nodes. Moody's algorithm utilizes a set of matrix formulas to derive the triad census. Batagelj and Mrvar [13] offer a triad census algorithm with a computational complexity of $O(m)$, where $m$ equals the number of edges. For the large, sparse graphs in which we are mainly interested, Batagelj and Mrvar's triad census algorithm has the potential to yield very good computational performance. Thus, we selected this algorithm to parallelize.

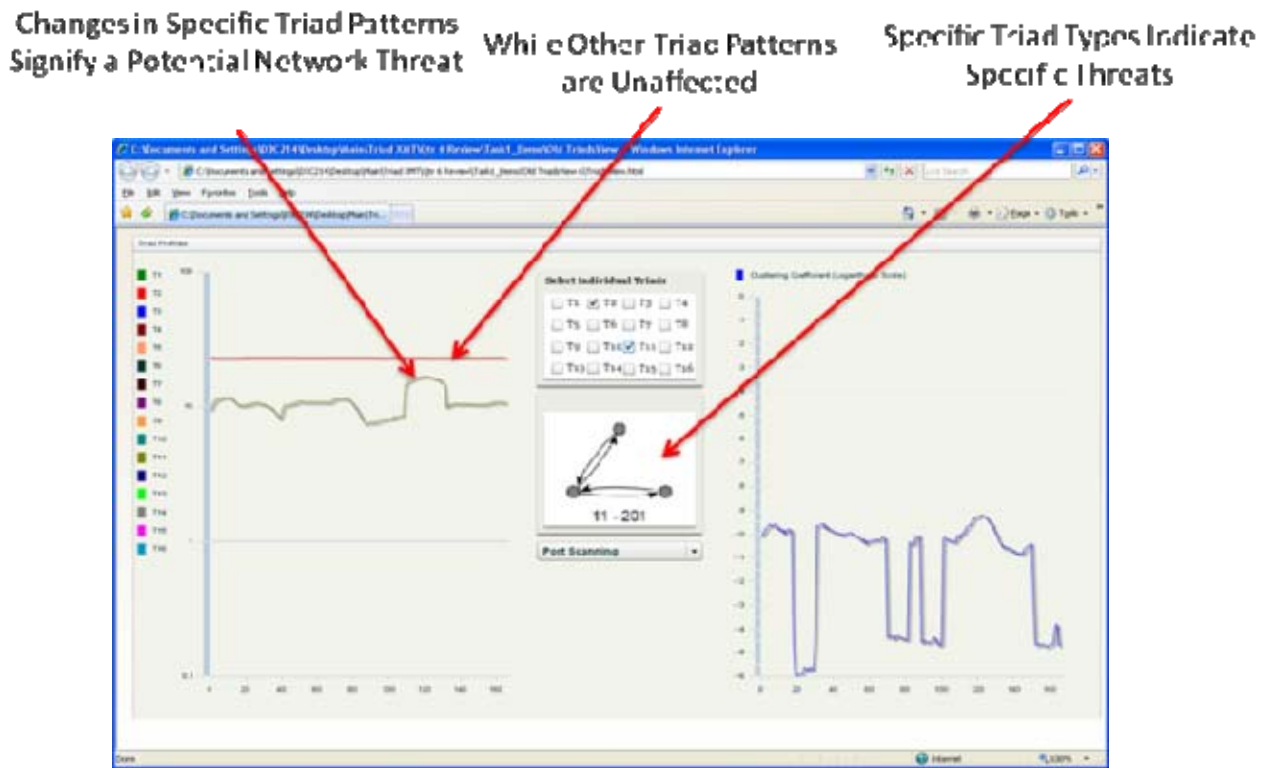

**Figure 4. Computer network security threat and anomaly monitoring tool.**

Batagelj and Mrvar's triad census algorithm is presented in Fig. 5. Inputs to the algorithm are a directed graph $G = (V, E)$, where $V$ is the set of vertices and $E \subseteq V \times V$ is the set of directed edges, and an array of neighbor lists $N$, which may be indexed by a node number. The output *Census* array collects the triad census. Given nodes $u$ and $v$ and the relation $A$, $uAv$ is true should an edge exist from $u$ to $v$ in $E$ or $\{u,v\} \in E$. Given nodes $u$ and $v$ and the relation $\hat{A}$, $u\hat{A}v$ is true should $u$ be a neighbor of $v$.

The algorithm works by following existing edges in the network. In step 2, $u$ is assigned to every vertex in $V$. In step 2.1, $v$ is assigned to every neighbor of $u$ that has a larger value than $u$. In step 2.1.4, w is assigned to each node of $S$, which is the union of the neighbors of $u$ and $v$. $u$, $v$, and $w$ make up the nodes of the triad that is currently being processed. In the inner processing of the algorithm, $u$ will always have the smallest value among $u$, $v$, and $w$.

The algorithm computes three different types of triads: null, dyadic, and connected. *Dyadic* triads have edges between two of three vertices. Given a pair of connected nodes, we compute the number of dyadic triads arising from the connected pair as $n - |S| - 2$ as shown in step 2.1.3. If a third node connects to either node of the connected pair, we then have a *connected* triad, where each node of the triad is connected to at least one edge. In step 2.1.4, we examine every node in *S* as the possible third node to the current triad. Here, we wish to avoid counting the same three nodes through different iterations of the code by only counting the canonical selection from *(u, v, w)* and *(u, w, v)*. If $u < w < v$ and *uÂw*, then *(u, w, v)* had already been considered in the algorithm. However, if ¬*uÂw*, then *(u, w, v)* is the canonical selection. In step 2.1.4.1, given the nodes of a connected triad, the *IsoTricode* function identifies the triad's isomorphic state via a lookup table, which may then be used to index into the *Census* array. In step 5, the number of *null* triads is computed as $(1/6)n(n-1)(n-2) - sum$, which is the total number of possible triads minus the number of triads with at least one edge.

```
INPUT: G = (V, E), N – array of neighbor lists
OUTPUT: Census array with frequencies of triadic types
1        for i := 1 to 16 do Census[i] := 0; \\ initialize census
2        for each u ∈ V do begin
2.1        for each v ∈ N[u] do if u < v then begin
2.1.1        S := N[u] ∪ N[v];
2.1.2        if uAv ∧ vAu then TriType := 3 else TriType := 2;
2.1.3        Census[TriType] := Census[TriType] + n − |S| − 2;
2.1.4        for each w ∈ S do if v < w ∨
                         (u < w ∧ w < v ∧ ¬uÂw) then begin
2.1.4.1        TriType := IsoTricode(u,v,w);
2.1.4.2        Census[TriType] := Census[TriType] + 1;
             end;
           end;
         end;
3        sum := 0;
4        for i := 2 to 16 do sum := sum + Census[i];
5        Census[1] := (1/6)n(n − 1)(n − 2) − sum;
```

**Figure 5. Batagelj and Mrvar's subquadratic triad census algorithm.**

The algorithm follows the edges of a graph to identify its set of triads, and thus, has a computational complexity of $O(m)$, where $m$ equals the number of edges. For sparse graphs, $m = O(k(n)*n)$, where $k(n) << n$. Since most large networks are sparse, the algorithm has a subquadratic computational complexity for the majority of large graphs.

## 5. Graph data sets

To evaluate our parallel triad census algorithm, we identified and collected three large, real-world, scale-free graphs. As shown in the charts of Fig. 6, the distribution of the number of outdegree connections per node for each graph follows a power law distribution. The three graphs are as follows:

- US patent datasets are available from the National Bureau of Economic Research [14]. Patent citation data may be organized as a citation network where nodes represent patents and directed edges identify citations to other patents. We downloaded a citation network dataset consisting of 37.8 million nodes and 16.5 million edges. The network carries a power law exponent of 3.126 for its outdegree distribution.
- Orkut [15] is a social networking site that allows users to organize into virtual communities, publish profiles, upload and share pictures, video clips, and other user-generated content, and to link to one another's homepages. Social network datasets consisting of users as nodes and links between user homepages as edges are available from Orkut.com. We downloaded an Orkut dataset consisting of 3.1 million nodes and 234.4 million edges. The graph carries a power law exponent of 2.127 for its outdegree distribution.
- A Web graph collects a portion of the World Wide Web, where nodes represent Webpages and edges identify hyperlinks to other Webpages. The Laboratory of Web Algorithmics (LAW) at the University of Milano [16] provides Web graph datasets that are generated from Web crawls of the .uk domain. The Web graph we retrieved consists of 105.2 million nodes and 2.5 billion edges, and carries a power law exponent of 1.516 for its outdegree distribution.

## 6. Code optimizations for shared memory architectures

We implemented the parallel triad census algorithm using a compact data structure where graph nodes are stored as the elements of an array as shown in Fig. 7. The collective set of edges for all the nodes are stored in a single array, whose memory is allocated only once. Each node points to the starting location in the edge array where its edges are populated. As shown, the two lower bits of the integer used to store the edge node id are reserved to encode edge direction, where "01" would indicate a unidirectional edge from current to neighbor node, "10" would indicate a unidirectional edge from neighbor to current node, and "11" would indicate a bidirectional edge between the current and neighbor node. In effect, we implemented a compressed sparse row data structure with two bits reserved to indicate edge direction.

The edge count for each node is stored in the node's data structure. The set of edges for a particular node is sorted in the edge subarray to enable fast edge searching using binary search. Since the count of edges per individual node is maintained, this compact data structure enables the Cray XMT compiler to parallelize the control loops that traverse the edges of a node (steps 2 and 2.1 of Fig. 5). The parallelization was confirmed using the Cray XMT compiler analysis tool (Canal). The compact data structure also enabled good spatial locality of the data in shared memory.

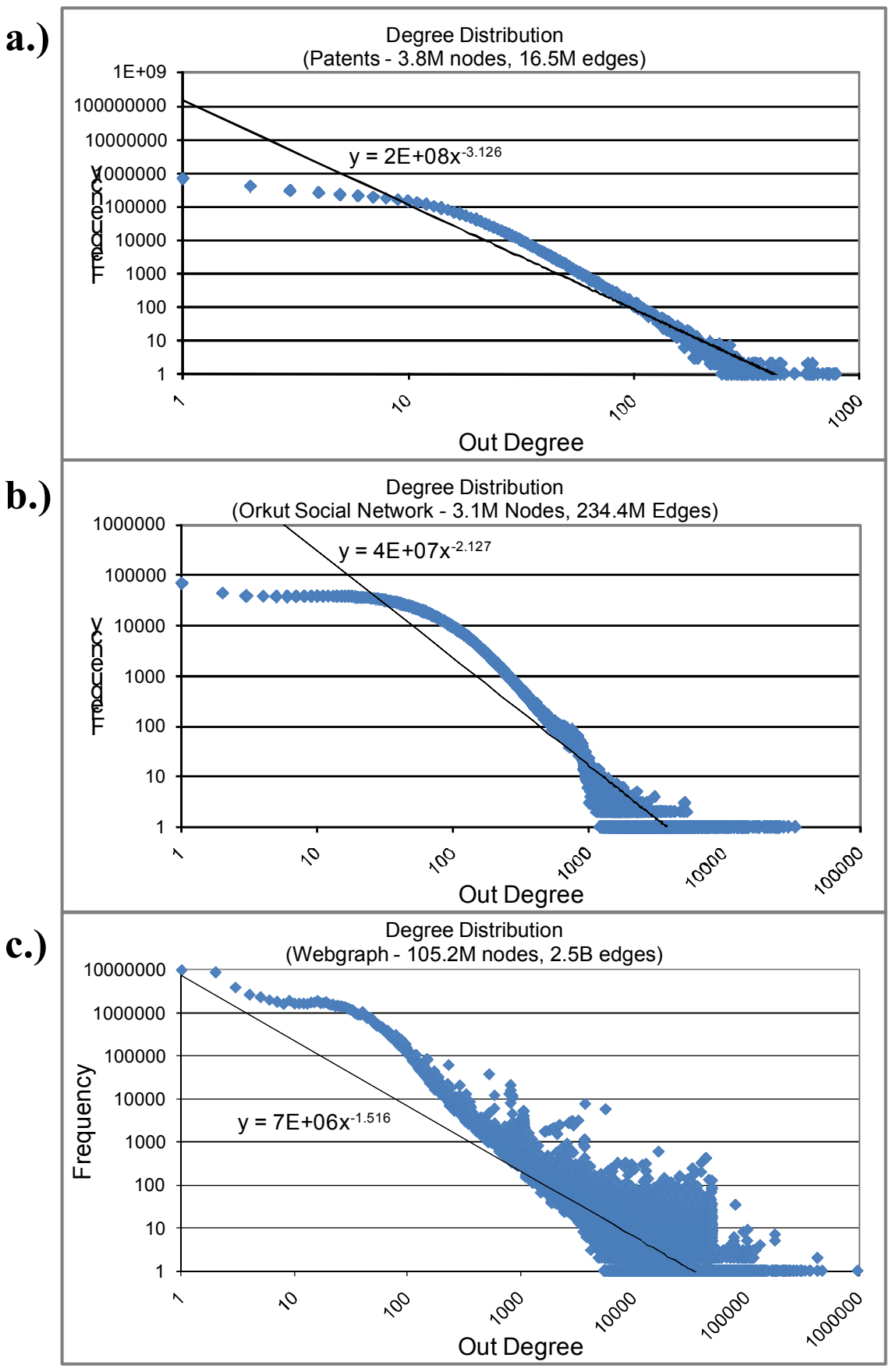

**Figure 6. Outdegree edge distribution charts for a.) patents network, b.) Orkut network, and c.) Webgraph of .uk domain.**

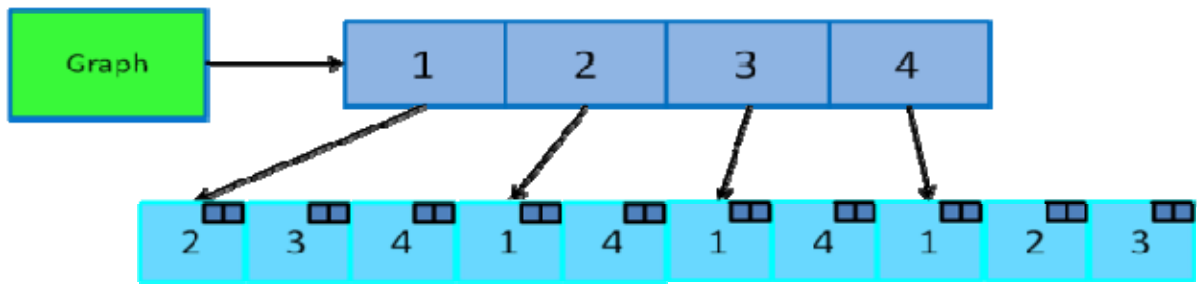

**Figure 7. Compact data structure representation for storing graph in parallel triad census algorithm.**

Another optimization focused on the manner in which the nodes of a triad were identified and processed. As shown in the pseudocode of Fig. 5, the first node *u* of the triad is one the vertices of the graph ($u \in V$), the second node *v* is one of the neighbors of the current vertex ($v \in \hat{A}(u)$), and the third node *w* is a vertex in the union of the neighbors of *u* and *v* with the exclusion of *v* and *u* ($w \in \hat{A}(u) \cup \hat{A}(v) \setminus \{v, u\}$). In previous versions of the triad census algorithm, we explicitly generated the union set *S*, and iterated through its elements to assign the third node of a triad. In the current version, however, we maintain two pointers that traverse the sorted neighbor arrays of *u* and *v*. As shown in Fig. 8, we traverse the two sorted arrays in numeric order by setting *w* to and incrementing the pointer referencing the lower value. In the case where the arrays have a common neighbor, both pointers are incremented after setting *w*.

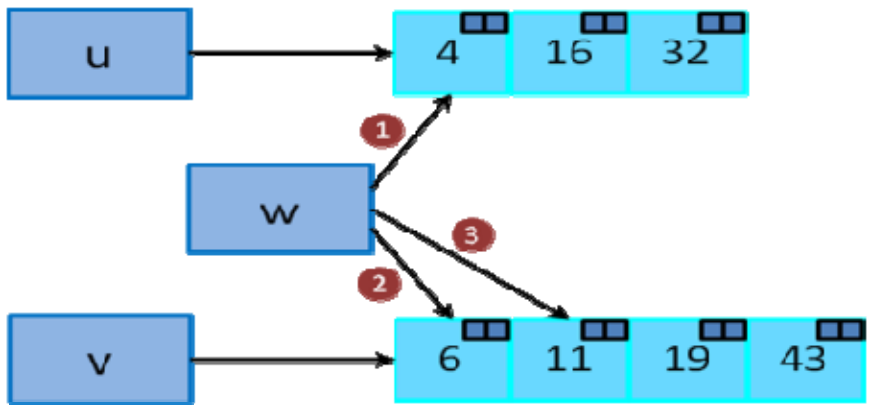

**Figure 8. *u*, *v*, and *w* point to the three nodes of the current triad. *w* traverses down the edge lists of *u* and *v* in numeric order based on the numbers in both lists. Circled numbers convey traversal order.**

The optimized data structure and neighbor array traversal strategy allow the code to maintain pointers to the relevant neighbor data, where it can quickly decipher edge directions using bit operations. This enables *in situ* identification of the triad pattern, where *w* pointing to a *u* neighbor identifies a *u-w* edge, *w* pointing to a *v* neighbor identifies a *v-w* edge, and *w* pointing to an equivalent *u* and *v* neighbor identifies both *u-w* and *v-w* edges. In all cases, the edge type is decoded from the two bits of the *w* neighbor.

One place for potential memory hotspotting in the triad census algorithm is the triad census vector. As triads of the graph are continuously identified using the algorithm, elements of the single 16-element triad census vector are constantly incremented and may act as points of continual contention. To alleviate this potential hotspotting, we implemented 64 local triad census vectors in the parallel triad census algorithm. Different iterations through the two outer *u* and *v* loops of the triad computation (steps 2 and 2.1 of Fig. 5) would increment different local triad census vectors based on a hash function that would accept *u* and *v* as a concatenated string and return a value inclusively between 0 and the number of triad censuses – 1. The hash function return values are uniformly distributed across the range of possible values. Once all triads have been identified, the 64 local triad census vectors are summed into a single final triad census vector.

As a preliminary test, we executed our parallel triad census algorithm on the Orkut network using 8 processors on the PNNL Cray XMT and collected overall CPU utilization rates at 10 second intervals. As shown in Fig. 9, the compact data structure version achieved a relatively consistent 60-70% CPU utilization rate over the course of execution after an initialization phase. In other XMT studies [17], developers found that the executions of most "well-tuned" XMT applications would peak at approximately 30% CPU utilization and have never experienced XMT applications executing at above 50% CPU utilization. Thus, the 60-70% CPU utilization rate for the compact data structure version of the triad census algorithm is considered to be very high for Cray XMT applications. The increased CPU utilization is an indicator that the exposed and exploited parallelism introduced opportunities for the compiler to favorably increase the ratio of register operations versus memory operations.

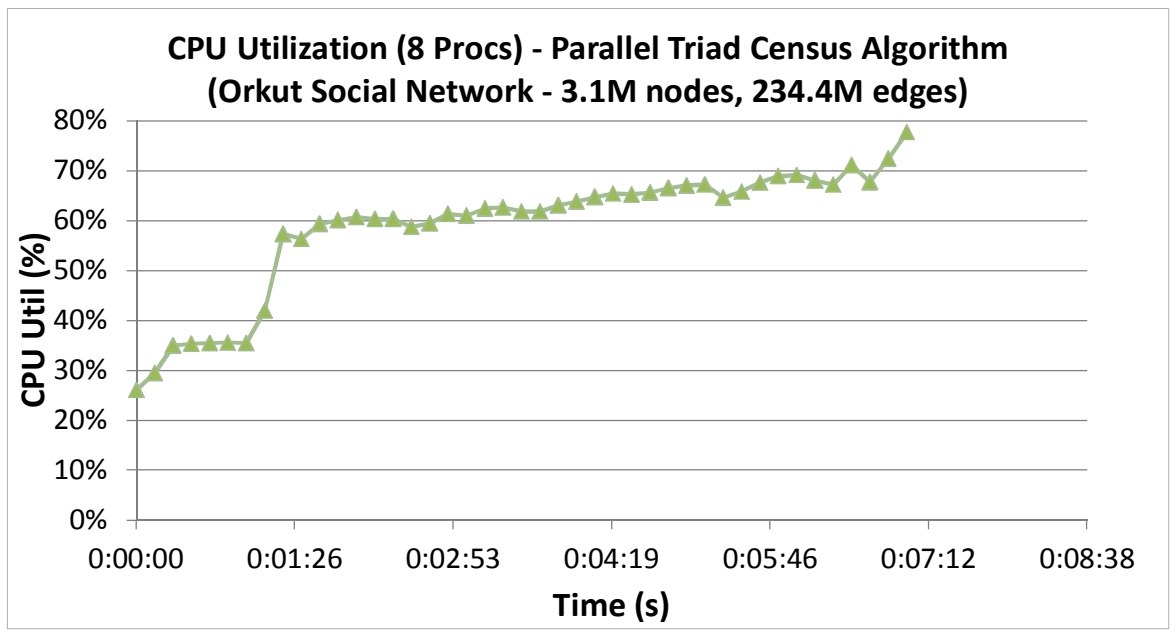

**Figure 9. CPU utilization rates for parallel triad census algorithm running Orkut network on 8 processors.**

# 7. Evaluating locality-agnostic triad algorithm across shared memory platforms

Using the Cray XMT, we have walked through the various algorithmic and data structure improvements that had been applied to the triad census algorithm. The improvements introduced are universally applicable for shared memory machines. We did not consider data partitioning and prefetching techniques for our problem formulation because the graph algorithms we are exploring prevent static and dynamic prediction of data access patterns. Yet, we did introduce private code tables and histograms to reduce contention induced by concurrent thread execution.

To investigate the triad census algorithm on an alternative shared memory machine, we ported the latest version of the algorithm to OpenMP. Atomic operations supported in memory by the XMT were mapped to OpenMP "atomic" pragmas. More so than the XMT, a HP Superdome would benefit from the algorithm's private data structures that migrate up the memory hierarchy by reducing access latency in addition to contention avoidance. Data compression like the embedded edge direction encoding would also contribute to memory bandwidth savings.

The HP Superdome at our disposal is a two cabinet, SD64 SX2000CEC with 8 cells per cabinet, each with 4 sockets, equipped with 1.6GHz dual-core Itaniums (Montecito) with 18MB cache. A crossbar hierarchy, sharing 256GB of interleaved memory, interconnects the cells. The total number of thread contexts supported in hardware is 256. The OpenMP C-compiler is Intel's version 10.1.

Initial analysis of the Superdome triad codes revealed that the Superdome compiler, as opposed to the XMT's, was not able to collapse the imperfectly nested loop over the graph's vertices and edges, and thus, yielded an unbalanced workload for initial sample graphs. After manually transforming the loops to produce a manhatten collapse, we were able to achieve a much improved balanced workload.

In addition, we investigated a massive multi-core NUMA machine with a total of 48 cores. Four 2.3GHz Opteron 6176SE 12-core processors (dual "Istanbul" 6 core die Magny-Cours), with 64kB+64kB L1 cache, 6MB L2 cache and 12MB L3 cache each, are linked by a ccNUMA 4xHT3 (4x6.4GT/s) interconnect and attached to 64GB DDR3 1.3GHz memory. We applied the same manual manhatten collapse to the NUMA codes after OpenMP compilation.

For our experimental discussion, we equate a core in the NUMA machine to a processor in a distributed memory machine (DMM) such as the XMT or Superdome. Our experimental results show only the best performing scheduling policy for each machine. For NUMA and Superdome, the best policy was a "dynamic" scheduling policy. Surprisingly, the "guided" policy severely underperformed.

Figs. 10a and 10b provide a glimpse of how the machines perform under limited concurrency opportunities as exposed by the patents network. The concurrency limitations manifest themselves in a limited, collapsed iteration space in the outer two loops in combination with an uneven workload in the innermost loop of the triad census algorithm. For a small number of cores, NUMA's overprovisioned memory bandwidth in combination with on-node, low-latency memory provides an architectural performance advantage the DMM architectures cannot match. Only at the point where all the NUMA's cores contribute to memory accesses is the XMT capable of overtaking the NUMA machine. As shown in Fig. 10a, the execution times cross at 36 processors for the NUMA and XMT machines. Performance degradation also begins at 36 processors for the NUMA machine before the 48 physical core limit is reached.

For similar reasons to NUMA, the Superdome is able to outperform the XMT up to the cell size of 8 cores. Crossing the cell boundary proves detrimental to the Superdome, whereas the XMT's fine-grain parallelism still manages to extract substantial performance gains up to 32 processors before slowly leveling off. In all experiments performed on the Superdome, scheduling strategies translate more pronouncedly into performance gains at architectural boundaries (cell, cabinet), with a "dynamic" scheduling strategy usually performing best. The scaling performance of the XMT is superior to both the NUMA and Superdome platforms as shown in Fig 10b.

a.)

b.)

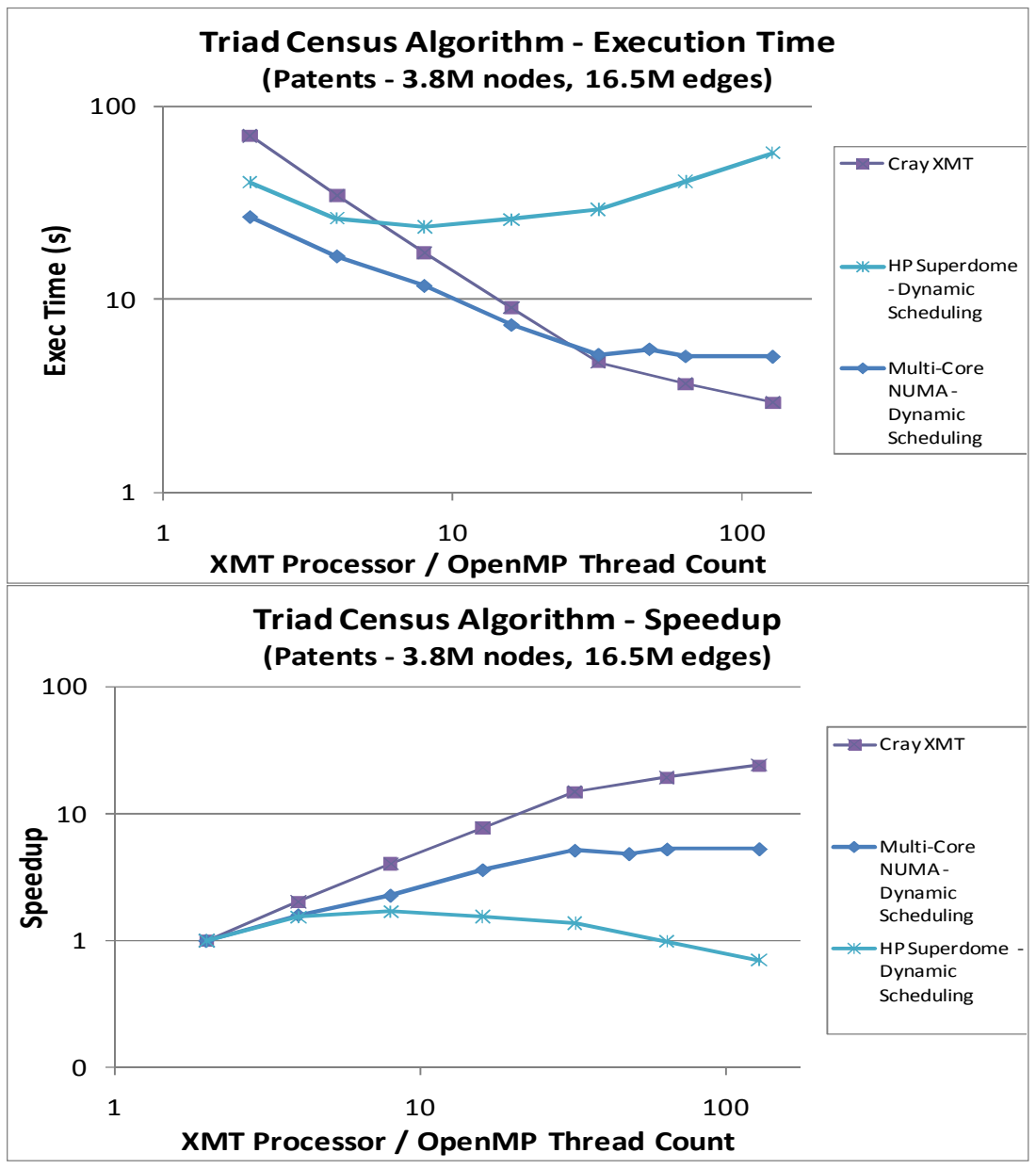

**Figure 10. Cray XMT, HP Superdome, and multi-core NUMA triad code performance running patents network at various core counts in terms of a.) execution times and b.) speedup rates.**

As shown in Figs. 11a and 11b, the performance of the NUMA and Superdome triad census codes drastically improve when processing the much larger Orkut network versus the patents network due to a largely increased outer loop iteration space that is able to better mask the unbalanced inner loop workload. Again, the NUMA manages to outperform the other machines at small core counts but this time it maintains its lead up to 64 virtual cores, in fact overprovisioning its physical 48 cores. As Fig. 12 illustrates in more detail, NUMA's performance degradation already becomes apparent at core counts in the 40s -- possibly attributed to memory oversubscription -- as its parallel efficiency starts to deteriorate. This behavior is in marked contrast to the XMT with an almost constant parallel efficiency.

Similarly, the Superdome achieves faster execution times than the Cray XMT until 64 cores for the corresponding dynamically scheduled code versions. Superdome's performance rate degradation at 64 cores can be attributed to a cabinet boundary crossing.

a.)

b.)

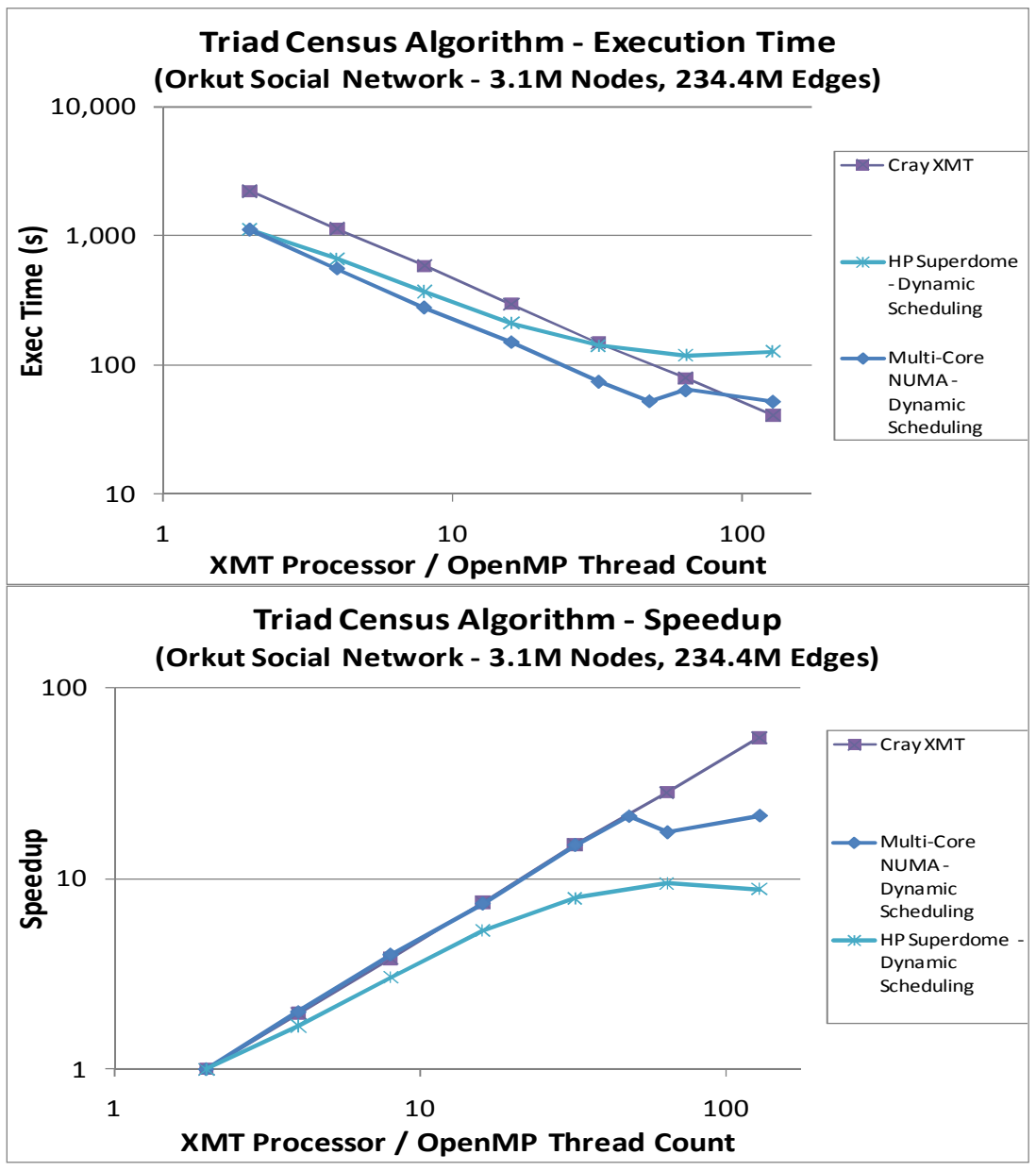

**Figure 11. Cray XMT, HP Superdome, and multi-core NUMA triad code performance running Orkut network at various core counts in terms of a.) execution times and b.) speedup rates.**

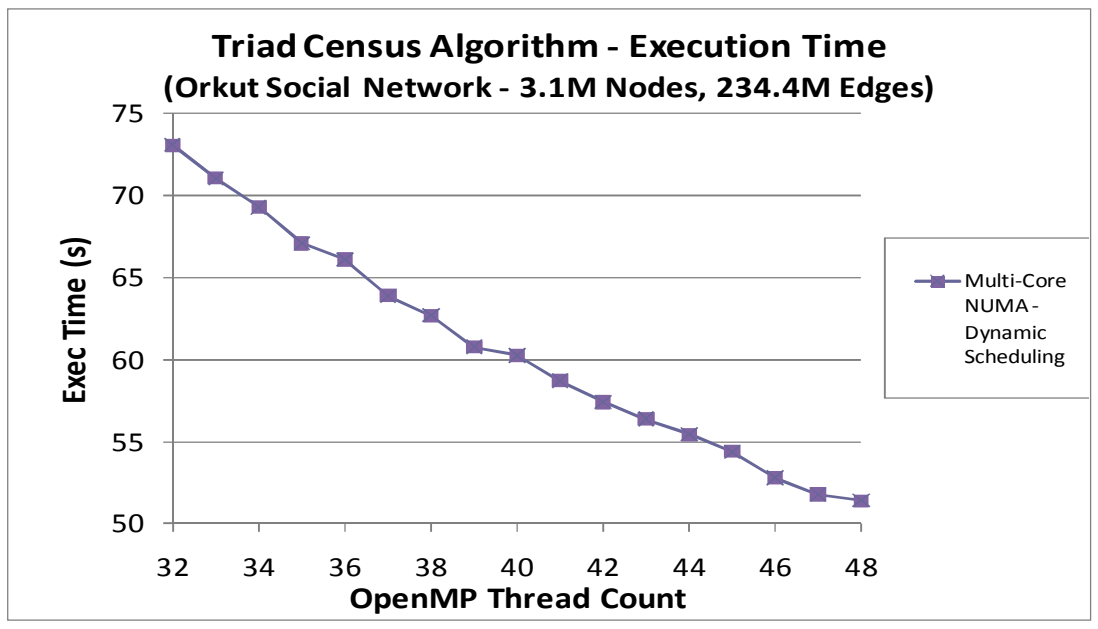

**Figure 12. Multi-core NUMA triad code execution time performance running Orkut network at core counts 32-48.**

As a final test, we executed the parallel triad census algorithm on a 512-processor XMT machine at Cray on the massive Webgraph network using 64 to 512 processors. Neither the NUMA nor Superdome machine was able to handle graphs of this size. As shown in Figs. 13a and 13b, the parallel triad census algorithm achieved good linear speedup rates from 64 to 512 processors. These results reinforce our observations from the "patents" and "orkut" experiments. Namely, massive concurrency opportunities are leveraged by the NUMA and the Superdome at smaller thread counts, leaving the XMT behind. Yet, as the thread count increases, their memory systems are overtaxed, which is a phenomenon we could not observe with the XMT.

Across the triad census experiments on all platforms, we discovered a recurring pattern: The XMT leaves some performance on the table at smaller processor counts but as processor counts increase, XMT's performance eventually outperforms the other machines' performances by leveraging its massive fine-grain multithreaded parallelism capabilities.

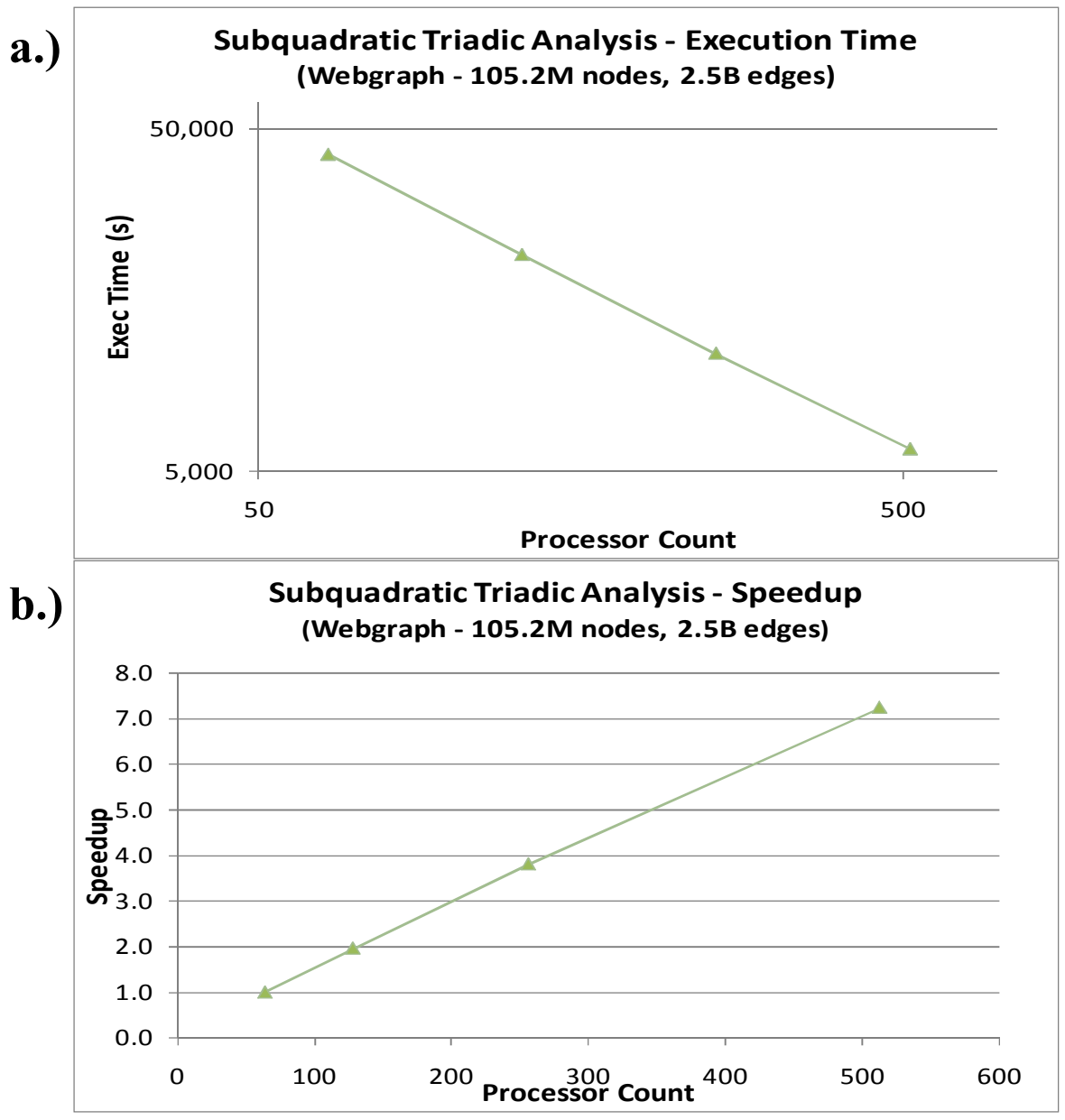

**Figure 13. Performance of parallel triad census algorithm running Webgraph network on the Cray XMT at various processor counts in terms of a.) execution times and b.) speedup rates.**

## 8. Conclusions

In this paper, we present a parallel triad census algorithm optimized to execute on shared memory machines. We describe the specific data structure and program logic optimizations we implemented to achieve a scalable algorithm. We also tested and evaluated the parallel triad census algorithm on different large-scale graphs and on different shared memory systems. A key contribution of this paper is to share our experiences and identify important development and optimization concepts and issues to consider when implementing and optimizing parallel graph algorithms on shared memory architectures.